\documentstyle{aipproc}

\def\ga{\gamma}
\def\de{\delta}
\def\ep{\epsilon}
\def\ve{\varepsilon}

\def\si{\sigma}

\def\ps{\psi}
\def\om{\omega}

\def\De{\Delta}

\def\fr#1#2{{{#1} \over {#2}}}

\def\half{{\textstyle{1\over 2}}}
\def\frac#1#2{{\textstyle{{#1}\over {#2}}}}

\def\lsim{\mathrel{\rlap{\lower4pt\hbox{\hskip1pt$\sim$}}
    \raise1pt\hbox{$<$}}}
\def\gsim{\mathrel{\rlap{\lower4pt\hbox{\hskip1pt$\sim$}}
    \raise1pt\hbox{$>$}}}
\def\sqr#1#2{{\vcenter{\vbox{\hrule height.#2pt
         \hbox{\vrule width.#2pt height#1pt \kern#1pt
         \vrule width.#2pt}
         \hrule height.#2pt}}}}

\newcommand{\beq}{\begin{equation}}
\newcommand{\eeq}{\end{equation}}
\newcommand{\bea}{\begin{eqnarray}}
\newcommand{\eea}{\end{eqnarray}}
\newcommand{\rf}[1]{(\ref{#1})}

\begin{document}
\title{Lorentz and CPT Tests in QED}

\author{Robert Bluhm}
\address{Physics Department, Colby College, Waterville, ME 04901, USA}

\maketitle

\begin{abstract}
A theoretical framework extending QED in the 
context of the standard model is used to
analyze a variety of Lorentz and CPT 
tests in atomic systems.
Experimental signatures of possible Lorentz
and CPT violation in these systems are investigated,
and bounds are discussed.
\end{abstract}

\section*{Introduction}

Lorentz symmetry and CPT appear to be exact fundamental symmetries of nature.
All of the known physical interactions seem to be invariant under continuous Lorentz
transformations consisting of boosts and rotations and under the
combined discrete symmetry CPT
formed from the product of charge conjugation C,
parity P,
and time reversal T.
These symmetries are connected by the CPT theorem
\cite{cptthm},
which states that under mild technical assumptions all local
relativistic field theories of point particles are symmetric under CPT.

Many of the shapest tests of Lorentz and CPT symmetry have
been performed in atomic systems where the predominant
interactions are described by quantum electrodynamics
\cite{pdg,cpt98}.
For example,
the Hughes-Drever type experiments 
\cite{cctests}
typically compare two
clocks or high-precision magnetometers involving different atomic species.
The best CPT tests for leptons and baryons cited by the
Particle Data Group are made by atomic physicists using Penning traps.
These experiments have obtained a bound on the $g$-factor 
difference for electrons and positrons given by 
\cite{epenning}
\beq
r_g^e = \fr {|g_{e^-} - g_{e^+}|} {g_{\rm avg}}
\lsim 2 \times 10^{-12}
\quad ,
\label{rge}
\eeq
while experiments with protons and antiprotons have obtained bounds on 
the difference in their charge-to-mass ratios given by
\cite{ppenning}
\beq
r_{q/m}^p = \fr{|(q_p/m_p)
- (q_{\overline{p}}/m_{\overline{p}})|} {(q/m)_{\rm avg}}
\lsim
9 \times 10^{-11}
\quad .
\label{rqmp}
\eeq
Similarly,
two proposed experiments at CERN intend to make high-precision
comparisons of the 1S-2S transitions in trapped hydrogen and antihydrogen
\cite{cernhbar}.
The 1S-2S transition is a forbidden transition that can only occur as
a two-photon transition.
It has a long lifetime and a small relative linewidth of approximately $10^{-15}$.
Atomic experimentalists believe that ultimately the line center might
be measured to 1 part in $10^3$ yielding a CPT bound
\beq
r_{\rm 1S-2S}^H = \fr{|\De \nu_{\rm 1S-2S} |} {\nu_{\rm 1S-2S}}
\lsim
\,\, 10^{-15} - 10^{-18}
\quad .
\label{r1s2s}
\eeq

The main focus of the work described here is to investigate
these and other tests of Lorentz and CPT symmetry performed in QED systems.
The general goals of this work have been to analyze the sensitivity
of QED systems to possible Lorentz and CPT violation,
to uncover possible new signals that can be tested in experiments,
and to estimate attainable experimental bounds in the context of a common 
framework that permits comparisons across different experiments.
To accomplish this,
we used the standard-model extension developed by
Kosteleck\'y and collaborators
\cite{ks,kp,ck}
as our theoretical framework.
This model permits a detailed investigation of Lorentz and CPT tests
in all particle sectors of the standard model.
Our analysis here focuses on the QED sector of the standard-model extension. 
This is presented in the following sections and is used
to examine five different QED systems:
experiments in Penning traps
\cite{bkr12},
clock-comparison tests
\cite{kl99},
experiments with hydrogen and antihydrogen
\cite{bkr3},
Lorentz and CPT tests with macroscopic spin-polarized materials
\cite{bk00}
and muon experiments
\cite{bkl00}.

As a result of our investigations,
several atomic experimental groups have recently reanalyzed existing data or taken
new data to obtain improved bounds on Lorentz and CPT violation.
Summaries of these bounds as well as estimates of bounds that can be attained in 
future experiments are presented in the following sections.

\section*{Standard-Model Extension}

Many different ideas for violation of Lorentz or CPT symmetry have been put
forward over the years since the proof of the CPT theorem.
A sampling of these includes the following:
nonlocal interactions  
\cite{carr68},
infinite component fields 
\cite{ot68},
a breakdown of quantum mechanics in gravity
\cite{hawk82},
Lorentz noninvariance at a fundamental level
\cite{np83},
spontaneous Lorentz violation 
\cite{ks},
and CPT violation in string theory 
\cite{kp,emn96}.

To explore the experimental consequences of possible Lorentz or CPT violation,
a common approach is to introduce phenomenological parameters.
Examples of this include the anisotropic inertial mass
parameters in the model of Cocconi and Salpeter
\cite{cs58},
the $\de$ parameter used in kaon physics
\cite{lw66},
and the TH$\ep\mu$ model which couples gravity and electromagnetism
\cite{ll73}.
Another approach is to introduce specific lagrangian terms
that violate Lorentz or CPT symmetry
\cite{cfj,cg}.
These approaches have the advantages that they are straightforward
and are largely model independent.
However,
they also have the disadvantages that their relation to experiments 
can be unclear and they can have limited predictive ability.
To make further progress,
one would want a consistent fundamental theory with CPT and Lorentz violation.
This would permit the calculation of phenomenological parameters
and the prediction of signals indicating symmetry violation.
No such realistic fundamental theory is known at this time.
However,
a candidate extension of the standard model incorporating
CPT and Lorentz violation does exist.

The standard-model extension of Kosteleck\'y and collaborators
is an effective theory based on the idea that Lorentz and CPT
symmetry can be spontaneously broken in the context of a more fundamental theory
\cite{ks}.
It is motivated in part from string theory
\cite{kp}.
The idea is to assume the existence of a fundamental theory
in which Lorentz and CPT symmetry hold exactly
but are spontaneously broken at low energy.
As in any theory with spontaneous symmetry breaking,
the symmetries become hidden at low energy.
The effective low-energy theory contains the standard model
as well as additional terms that could arise through the symmetry breaking process.
A viable realistic fundamental theory is not known at this time,
though higher dimensional theories such as string or M theory are promising candidates.  
A mechanism for spontaneous symmetry breaking
can be realized in string theory because
suitable Lorentz-tensor interactions can arise which destabilize
the vacuum and generate nonzero tensor vacuum expectation values.

Colladay and Kosteleck\'y have derived the most general extension
of the standard model that could arise from spontaneous Lorentz
symmetry breaking of a more fundamental theory,
maintains SU(3)$\times$SU(2)$\times$U(1) gauge invariance,
and is renormalizable
\cite{ck}.
They have shown that the theory maintains many of
the other usual properties of the standard model besides Lorentz and CPT symmetry,
such as electroweak breaking, energy-momentum conservation,
the spin-statistics connection, and observer Lorentz covariance.
Issues related to the stability and causality of the standard-model
extension are investigated in
Ref.\ \cite{kl}.
In addition to the atomic experiments described here,
the standard-model extension has been used to analyze
Lorentz and CPT tests with neutral mesons
\cite{mesontests,ckpv},
photon experiments
\cite{ck,cfj,jk99},
and baryogenesis
\cite{bckp}.

\section*{TESTS IN ATOMIC SYSTEMS}

To consider experiments in atomic systems it is sufficient
to restrict the standard-model extension to its QED sector.
The modified Dirac equation for a
four-component spinor field $\ps$
of mass $m_e$
and charge $q = -|e|$ in an electric potential $A^\mu$ is
\beq
\left( i \ga^\mu D_\mu - m_e - a_\mu \ga^\mu
- b_\mu \ga_5 \ga^\mu - \half H_{\mu \nu} \si^{\mu \nu}
+ i c_{\mu \nu} \ga^\mu D^\nu
+ i d_{\mu \nu} \ga_5 \ga^\mu D^\nu \right) \ps = 0
\quad .
\label{dirac}
\eeq
Natural units with $\hbar = c = 1$ are used,
and $i D_\mu \equiv i \partial_\mu - q A_\mu$.
The two terms involving the effective coupling constants
$a_\mu$ and $b_\mu$ violate CPT,
while the three terms involving
$H_{\mu \nu}$, $c_{\mu \nu}$, and $d_{\mu \nu}$
preserve CPT.
All five terms break Lorentz invariance.
Each particle sector in the standard model has its own set
of parameters which we distinguish using superscripts.
Since no Lorentz or CPT violation has been observed,
these parameters are assumed to be small.
A perturbative treatment in the context of relativistic
quantum mechanics can then be used.
In this approach,
all of the perturbations in conventional quantum electrodynamics
are identical for particles and antiparticles.
However,
the interaction hamiltonians 
including the effects of possible
Lorentz and CPT breaking are not the same.

\subsection*{Penning-Trap Experiments}

Comparisons of the $g$ factors and charge-to-mass ratios of
particles and antiparticles confined within a Penning trap
have yielded the CPT bounds in
Eqs.\ \rf{rge} and \rf{rqmp}.
These quantities are obtained through measurements of the
anomaly frequency $\om_a$ and the cyclotron frequency $\om_c$.
For example,
$g-2=2\om_a/\om_c$.
These frequencies can be measured to $\sim 10^{-9}$ thereby
determining $g$ to $\sim 10^{-12}$.

We have analyzed Penning-trap experiments with electrons and positrons
and with protons and antiprotons
\cite{bkr12}.
We find that to leading order in the Lorentz-violating parameters
there are corrections to the anomaly frequencies that are different
for particles and antiparticles,
while the cyclotron frequencies receive corrections that
are the same for particles and antiparticles.
Both frequencies have corrections that cause them to exhibit
sidereal time variations.
We also find that to leading order the $g$ factor has no corrections, 
and therefore the figure of merit $r_g^e \simeq 0$,
even though there is explicit CPT breaking.
Because of this,
we have proposed using as an alternative figure of merit
the relative relativistic energy shifts caused by Lorentz and CPT violation.
This is a definition that can be used in any experiment and is
consistent with neutral meson experiments, 
which use mass ratios.

Based on these observations,
we suggested looking for two signals of Lorentz and CPT violation:
one an instantaneous difference in anomaly frequencies for electrons and positrons, 
and the other sidereal-time variations in the anomaly frequency of electrons alone.
Dehmelt's group at the University of Washington has recently published two papers
with the results of these observations
\cite{dehmelt}.
In their first paper,
they reanalyzed existing data and obtained a figure of merit
$r^e_{\om_a} \lsim 1.2 \times 10^{-21}$ 
from a bound on the difference in the electron and positron anomaly frequencies.
In the second paper, 
they analyzed more recent data for the electron alone and obtained
a bound on sidereal time variations given by
$r^e_{\om_{a}^{-},\rm diurnal} \lsim 1.6 \times 10^{-21}$.
This corresponds to a bound on the combination of components 
$\tilde b_J^e  \equiv b_J^e - m d_{J0}^e - \half \ve_{JKL} H_{KL}^e$
defined with respect to a nonrotating coordinate system
\cite{kl99} 
given by
$|\tilde b_J^e| \lsim 5 \times 10^{-25} {\rm GeV}$.

Although no $g-2$ experiments have been made for protons or antiprotons,
there have been recent bounds obtained on Lorentz violation in comparisons
of cyclotron frequencies of antiprotons and $H^-$ ions confined in a Penning trap
\cite{ppenning}.
In this case the sensitivity is to the parameters $c^p_{\mu \nu}$, 
and the figure of merit $r^{H^-}_{\om_c} \lsim 10^{-25}$ was obtained.

\subsection*{Clock-Comparison Experiments}

The Hughes-Drever type experiments
are atomic clock-comparison tests of Lorentz invariance
\cite{cctests}.
These experiments look for relative changes between two ``clock''
frequencies as the Earth rotates.
The ``clock'' frequencies are typically atomic Zeeman or hyperfine transitions.
Kosteleck\'y and Lane
\cite{kl99}
have made an extensive analysis of these experiments
using the standard-model extension.
They have obtained approximate bounds on various combinations
of the Lorentz-violating parameters from the published results
of these experiments.
For example,
from the experiment of Berglund {\it et al.}
the bounds 
$\tilde b_J^p \sim 10^{-27}$ GeV,
$\tilde b_J^n \sim 10^{-30}$ GeV,
and $\tilde b_J^e \sim 10^{-27}$ GeV 
have been obtained,
respectively,
for the proton, neutron, and electron sectors.

Since certain assumptions about the nuclear configurations
must be made to extract some of these numerical bounds,
these bounds should be viewed as good to within one
or two orders of magnitude.
To obtain cleaner bounds it is necessary to consider
simpler atoms.
This is done in the following sections,
where it is shown that sharper bounds can be obtained for the
proton and electron sectors.
However,
the clock-comparison tests continue to provide the best bounds
for the neutron sector.
For example,
a recent experiment using a two-species noble-gas maser
is consistent with there being no Lorentz or CPT violation 
in the neutron sector at a level of $10^{-31}$ GeV
\cite{dualmaser}.
This is currently the sharpest test for the neutron.

\subsection*{Hydrogen-Antihydrogen Experiments}

We have investigated the proposed experiments at CERN which will
make high-precision spectroscopic measurements of the 1S-2S
transitions in hydrogen and antihydrogen
\cite{bkr3}.
We find that the magnetic field plays a crucial role
in the sensitivity of the 1S-2S transition to Lorentz and CPT breaking.
For example,
in free hydrogen in the absence of a magnetic field,
the 1S and 2S levels shift by an equal amount
at leading order in hydrogen and antihydrogen. 
Because of this,
there are no leading-order corrections to the 1S-2S transition
frequency in free H or $\bar {\rm H}$.
For hydrogen in a magnetic trap,
there are magnetic fields which mix the spin states in the
four hyperfine levels.
Since the Lorentz-violating couplings are spin-dependent,
some of the 1S and 2S levels acquire energy corrections that are not equal.
The transitions between these levels have leading-order sensitivity 
to Lorentz and CPT violation.
However,
these transitions are field-dependent,
making them prone to broadening in an inhomogeneous magnetic field.
To be sensitive to leading-order Lorentz and CPT violation in 1S-2S transitions, 
experiments would have to overcome the difficulties
associated with possible line broadening effects due to field inhomogeneities.

We have also considered measurements of the ground-state Zeeman
hyperfine transitions in hydrogen and antihydrogen
\cite{bkr3}.
We find that certain transitions in a hydrogen maser 
are sensitive to leading-order Lorentz-violating effects.
These measurements have now been made by the group of Walswoth 
at the Harvard-Smithsonian Center using a double-resonance technique
\cite{Hmaser}.
They have obtained bounds on the Lorentz-violation parameters
for the proton and electron.
The bound for the proton is $|\tilde b_J^p| \lsim 10^{-27}$ GeV.
This is an extremely clean bound and is currently the most stringent test
of Lorentz and CPT symmetry for the proton.

\subsection*{Spin-Polarized Matter}

Experiments at the University of Washington with a spin-polarized
torsion pendulum 
\cite{eotwash}
are able to achieve very high sensitivity to
Lorentz violation due to the combined effect of a large number
of aligned electron spins
\cite{bk00}.
The experiment uses stacked toroidal magnets that have a net
electron spin $S \simeq 8 \times 10^{22}$,
but which have a negligible magnetic field.
The apparatus is suspended on a turntable and a time-varying
harmonic signal is sought.
Our analysis shows that in addition to a signal with the
period of the rotating turntable,
the effects of Lorentz and CPT violation would induce additional
time variations with a sidereal period caused by Earth's rotation.
The University of Washington group has analyzed their data
and have obtained a bound on the electron parameters
equal to $|\tilde b_J^e| \lsim 1.4 \times 10^{-28}$ GeV
\cite{heckel}.
This is currently the best Lorentz and CPT bound for the electron.

\subsection*{Muon Experiments}

Despite the high precision of recent Lorentz and CPT
tests for neutrons, protons, and electrons,
it is important to keep in mind that particle
sectors in the standard-model extension can be independent of each
other and should separately be tested.
The situation is similar to CP tests where violation is observed
only in the neutral meson sector and not in the lepton or baryon sectors.
A thorough investigation of Lorentz and CPT symmetry should therefore
probe as many possible particle sectors as possible.
For this reason,
we also consider muon experiments,
which involve second-generation leptons.
We find that there are several different types of experiments that
are sensitive to Lorentz and CPT.
We have examined both experiments in muonium
\cite{muonium99}
and $g-2$ experiments with muons being conducted at Brookhaven
\cite{muong99}.

Our results are that experiments measuring the frequencies
of ground-state Zeeman hyperfine transitions
in muonium in a strong magnetic field are sensitive
to Lorentz and CPT violation.
If bounds on sidereal time variations are obtained at
the 100 Hz level,
then the Lorentz-violation parameter for the muon $\tilde b_J^\mu$
can be bounded at the level of $| \tilde b^\mu_J| \le 5 \times 10^{-22}$ GeV.
We also find that in relativistic $g-2$ experiments using positive muons
with ``magic'' boost parameter $\de = 29.3$, 
bounds on Lorentz-violation parameters are possible at
a level of $10^{-25}$ GeV.
Experiments looking for sidereal time variations in
the muon anomaly frequency would yield stringent new Lorentz and CPT bounds.

\section*{SUMMARY AND CONCLUSIONS}

By using an extension of QED incorporating Lorentz and CPT violation, 
we have been able to analyze a variety of atomic experiments.
We have shown that low-energy experiments in QED systems are
sensitive to suppression factors associated with the Planck scale. 
We also find that experiments traditionally considered Lorentz tests 
are sensitive to CPT, and vice versa.
The atomic experiments considered here complement those in particle physics and
together they are able to test the robustness of the standard model
to increasing levels of precision.

\begin{center}
{\large \bf ACKNOWLEDGMENTS}
\end{center}

I would like to acknowledge my collaborators
Alan Kosteleck\'y, Charles Lane, and Neil Russell.
This work was supported in part
by the National Science Foundation
under grant number PHY-9801869.

\end{document}